\documentclass[a4paper, 11pt]{article}
\usepackage[utf8]{inputenc}
\usepackage{indentfirst}
\usepackage{geometry}
\usepackage{textcomp}
\usepackage{amsmath}
\usepackage{amssymb}
\usepackage{graphicx}
\usepackage{color,xcolor}
\usepackage{mathrsfs}
\usepackage{caption}
\usepackage{dsfont}
\usepackage[colorlinks=true, allcolors=blue]{hyperref}
\usepackage{tcolorbox}
\usepackage{tikz}
\usepackage{tikz-cd}
\usepackage{physics}
\usepackage{diagbox}
\usepackage{listings}
\usepackage{appendix}
\usepackage{array}
\usepackage{subfig}

\usepackage{cite}

\catcode`\>=\active \def>{
\fontencoding{T1}\selectfont\symbol{62}\fontencoding{\encodingdefault}}

\usepackage{amsmath}
\numberwithin{equation}{section}
\usepackage[normalem]{ulem}
\usepackage{color}
\definecolor{darkgreen}{RGB}{40,150,60}

\def \hs{\hspace}

\def\be{\begin{equation}}
\def\ee{\end{equation}}
\def\bag{\begin{aligned}}
	\def\eag{\end{aligned}}
\def\bea{\begin{eqnarray}}
\def\eea{\end{eqnarray}}
\def\ba{\begin{array}}
	\def\ea{\end{array}}

\title{Carrollian superstring in the flipped vacuum}
\author{Bin Chen$^{1,2,3}$, and Zezhou Hu$^2$}
\date{\today}

\begin{document}

\maketitle
\begin{center}
	{\it
		$^{1}$Institute of Fundamental Physics and Quantum Technology,\\ \&  School of Physical Science and Technology, \\ Ningbo University, Ningbo, Zhejiang 315211, China\\\vspace{2mm}
        $^2$School of Physics, Peking University, No.5 Yiheyuan Rd, Beijing 100871, P.~R.~China\\
		\vspace{2mm}
		$^{3}$Center for High Energy Physics, Peking University, No.5 Yiheyuan Rd, Beijing 100871, P.~R.~China\\
	}
	\vspace{10mm}
\end{center}

\begin{abstract}
    \vspace{5mm}
In this work, we study the spectrum of the Carrollian superstring in the flipped vacuum (the highest-weight representation) in detail. The target spacetime of the Carrollian string has a generalized Carrollian symmetry, and it is composed of both Poincar\'e directions and Carrollian directions. We explicitly show that two homogeneous Carrollian superstring theories, one with compactified Poincar\'e direction and the other with compactified Carrollian direction, share the same homogenous super-$\mathfrak{bms}_3$ symmetry, and their zero modes can be identified under the usual T duality transformation. Moreover, we investigate the spectrum of a general Carrollian superstring in the flipped vacuum. As the string can still have nonvanishing winding along the spatial direction in the infinite radius limit, the spectrum of the Carrollian strings in the flat background is no longer truncated. We furthermore construct the vertex operators of gravitons and discuss their scattering amplitudes. We find that the form of 3-point amplitudes differs from that of the usual tensile superstrings only by a simple function, which reflects the ultralocal nature of Carrollian physics.


\end{abstract}

\vfill{\footnotesize E-mail: chenbin1@nbu.edu.cn, z.z.hu@pku.edu.cn}

\newpage

\section{Introduction}

The tensionless limit of string theory presents rich symmetries of string theory at high energy\cite{Gross:1987ar, Gross:1987kza, Gross:1988ue}. Intuitively, under the tensionless limit $\alpha'\rightarrow \infty$,  all massive string excitations with $m^2\sim\frac{1}{\alpha'}$ become massless, and these massless modes have rich symmetries and are expected to compose a higher spin theory. However, as the consistent higher spin theory with interaction can only be defined in spacetimes with a nonzero cosmological constant\cite{Fradkin:1987ks, Vasiliev:2003cph}, the naive tensionless limit of flat string cannot be a higher spin theory. 

There is another way to investigate the tensionless string from a world sheet point of view. It was first discussed in \cite{Schild:1976vq}. In \cite{Isberg:1993av}, the world sheet symmetry of the tensionless string was studied in detail. The symmetry is generated by 2D Carrollian conformal algebra\cite{Bagchi:2013bga}, which is isomorphic to the $\mathfrak{bms}_3$ algebra. The quantum aspects of the tensionless string have been studied to some extent \cite{Isberg:1993av, Bagchi:2015nca}. One subtle issue in the tensionless string is its spectrum, even in a flat spacetime background, since there are three different options on vacuum\cite{Bagchi:2020fpr, Bagchi:2021rfw}:  oscillator vacuum, induced vacuum, and flipped vacuum. The oscillator vacuum is the weirdest one because one cannot impose traditional physical conditions on it. The induced vacuum sounds natural as it is the direct tensionless (ultrarelativistic) limit of the original tensile string theory. And the flipped vacuum can be obtained from the limit of the ``flipped" string, where its antiholomorphic part is reversed as $\bar{L}_{n}\rightarrow -\bar{L}_{-n}$. However, it is difficult to establish the state-operator correspondence, which is crucial for string theory, in the oscillator vacuum or the induced vacuum. The only successful realization of state-operator correspondence is based on the flipped vacuum (the highest weight representation)\cite{Hao:2021urq, Chen:2023naw}. But the spectrum built on the flipped vacuum is truncated and obviously cannot match the one in a higher spin theory\footnote{For the discussions of the quantization on different vacua and state-operator correspondence in higher-dimensional Carrollian conformal scalar, see \cite{Chen:2024voz}.}. 

Very recently, it was found that the tensionless string has an interesting relation to the Carrollian string\cite{Gomis:2023eav, Blair:2023noj}. The Carrollian string has a generalized Carrollian symmetry in its target spacetime\cite{Gomis:2023eav}, besides the same BMS$_3$ world sheet symmetry as the tensionless string. The target spacetime of the Carrollian superstring may have $10-q$ Poincar\'e directions and $q$ Carrollian directions. The tensionless string theory could be regarded as a special case of Carrollian string theory\footnote{Notice that there are two kinds of constructions on Carrollian strings, the electric type \cite{Bagchi:2024rje} and the magnetic type \cite{Bagchi:2023cfp}. Our construction in this work and that in \cite{Gomis:2023eav} belong to the electric type. In our language, the electric Carrollian string in \cite{Bagchi:2024rje} has two Poincar\'e directions and two Carrollian directions, see Eq.(4.5) in \cite{Bagchi:2024rje}. The world sheet of the electric type is a two-dimensional (2D) Bondi-Metzner-Sachs (BMS) field theory (see section 3.2.2 in \cite{Bagchi:2024rje}), while that of the magnetic type is not.} with $q=0$, and the conventional Carollian string is the case with $q=9$. At the level of the string sigma model, it was shown that T duality exists among the Carrollian superstring theories. Turning on world sheet SUSY, things become more complicated \cite{Bagchi:2016yyf, Bagchi:2017cte}. We will see that only the homogeneous superstrings are sensible and exhibit spacetime supersymmetry in section \ref{secSpectrum} of our paper. There, the supersymmetry (with homogeneous fermions) must be introduced, because all the excitations satisfying the physical constraints have zero norms in the bosonic string and inhomogeneous superstring case, some discussions can also be found in \cite{Chen:2023esw}.

In this work, we would like to investigate the Carrollian superstring more carefully, especially its spectrum and the associated vertex operators.  As the theory has the same $\text{BMS}_3$ symmetry, we have to face the problem of selecting the vacuum.  We will work in the flipped vacuum. We first consider the case with compactified spatial directions in order to check T duality explicitly. As the actions of the Carrolian directions are quite different from those of the Poincar\'e directions, the T duality in the spectra is not as simple as the one in standard string theory.  We will explicitly show that these superstring theories all have the same symmetry generated by homogeneous super-$\mathfrak{bms}_3$ algebra \cite{Bagchi:2018wsn}\footnote{The anomalous terms $A_L(m)$ and $A_M(m)$ and more details can be found in \cite{Chen:2023esw}.},
\be\label{bmsalgebra}
\bag
&[L_m,L_n]=(m-n)L_{m+n}+\delta_{m+n}A_L(m)\,,\\
&[L_m,M_n]=(m-n)M_{m+n}+\delta_{m+n}A_M(m)\,,\\
&[L_n,H_r]=(\frac{n}{2}-r)H_{n+r}\,,\quad \{H_r,H_s\}=2M_{r+s}\,,\\
&[L_n,\tilde{H}_r]=(\frac{n}{2}-r)\tilde{H}_{n+r}\,,\quad \{\tilde{H}_r,\tilde{H}_s\}=2M_{r+s}\,.
\eag
\ee
Secondly, we discuss the spectrum of the Carrollian superstring in a flat background by considering the infinite-radius limit of compactified spatial circles. It presents a few novel features: the winding number can still be nonzero, unlike in the tensile string theory\footnote{Similar discussion appears in the paper \cite{Banerjee:2023ekd}, which considered bosonic tensionless string theory compactified on circle and on torus.}; the spectrum is no longer truncated except for the special case $q=9$, which corresponds to the conventional Carrollian string. For the conventional Carrollian string, its spectrum is truncated to that of a conventional Carrollian gravity with supersymmetry. On the other hand, its momentum in the spatial directions is vanishing, which matches the ultralocal property of a conventional Carrollian field theory. Finally, we construct the vertex operators of the gravitons and study their amplitudes. We explicitly calculate the 3-point amplitudes of gravitons and discuss their implications for Carrollian gravity. 


We hope our study may shed new light on Carrollian physics. Carrollian symmetry has many important applications, e.g., in fracton\cite{Figueroa-OFarrill:2023vbj}, in the horizon (Rindler) physics\cite{Donnay:2019jiz, Freidel:2022vjq}, and in ultrarelativistic physics. Carrollian conformal field theory based on the highest-weight representation is at the center of flat holography \cite{Donnay:2022wvx, Nguyen:2023vfz, Chen:2023naw}. Moreover, Carrollian gravity, as a counterpart to Newton-Cartan gravity, explores the extreme limits of spacetime dynamics where the speed of light approaches zero. For the recent developments on Carrollian gravity, see \cite{Dautcourt:1997hb, Bekaert:2015xua, Hartong:2015xda, Bergshoeff:2017btm, Hansen:2021fxi, Henneaux:2021yzg, Perez:2022jpr, Figueroa-OFarrill:2022mcy, Campoleoni:2022ebj, Concha:2022muu, Sengupta:2022rbd,  Bergshoeff:2023rkk, Ecker:2023uwm, Ciambelli:2023tzb, deBoer:2023fnj, Musaeus:2023oyp,  Tadros:2023teq, Tadros:2024qlo, March:2024zck, Grumiller:2024dql, Chen:2024how, Bergshoeff:2024ilz, Concha:2024tcu}. It is well-known that the low-energy effective action of string theory is Einstein's gravity. It is an interesting question to investigate if one can obtain the Carrollian gravity or some of its features \cite{deBoer:2023fnj, Campoleoni:2022ebj} from string theory.

The remaining parts of this paper are organized as follows. In section \ref{section2}, we first introduce the RNS formalism of the Carrollian superstring theory and then compare the Carrollian superstring theory with a spatial direction $X^a$ compactified with radius $R_\text{T}$ to that with a spatial direction $X^i$ compactified with radius $R_\text{C}$. We explicitly show that they are T dual to each other by 
\be
k_{\text{C}}=w_{\text{T}},\quad w_{\text{C}}=k_{\text{T}},\quad R_{\text{C}}=1/(2R_{\text{T}})\,.
\ee
We further discuss the spectrum for a Carrollian string theory in a flat background with and without the winding numbers.
In section \ref{section3}, using the techniques of 2d BMS field theory, we explore the vertex operators associated with the gravitational modes. Then we use them to calculate the scattering amplitudes of gravitons and discuss their implications for Carrollian gravity.
In section \ref{section4}, we conclude and discuss our results and their further implications. 
In the appendix \ref{appendixC}, we present supplementary material on the inhomogeneous super-$\mathfrak{bms}_3$ algebra.


\section{The Carrollian string theory}\label{section2}
The Carrollian string theory has a generalized Carrollian symmetry\cite{Gomis:2023eav} in its spacetime coordinates $X^\mu, \mu=(a, i)$. Especially, the action is invariant under the Carroll-like boost transformation
\be\label{genCarr}
\delta X^a=\Lambda^a_i X^i\,,\qquad\delta X^i=0\,,
\ee
with $a=0,\cdots,9-q$ and $i=10-q,\cdots,9$. We call the coordinates $X^i$ Carrollian spatial directions, which form a subspace admitting $SO(q)$ symmetry. The other coordinates $X^a$ form a subspace admitting a Poincar\'e symmetry $SO(1,9-q)$. In the literature, conventional  Carrollian symmetry is referred to as the one in the special case $q=9$. And the $q=0$ case returns to the tensionless string.

We will focus on $\mathcal{N}=2$ homogenous Carrollian superstrings because it has normalizable excitations and exhibit spacetime supersymmetry. In the Ramond-Neveu-Schwarz (RNS) formulation, the action for the Carrollian superstring is of the form
\be\label{actionCo}
\bag
S=&\frac{1}{2\pi}\int d^2 \sigma e [e^{\alpha}_0 e^{\beta}_0 \partial_\alpha X^a\partial_\beta X^b\eta_{a b}+i \eta_{a b}e^{\alpha}_0 (\psi^a\partial_\alpha\psi^b+\tilde{\psi}^a\partial_\alpha\tilde{\psi}^b)\\
&+(-e^{\alpha}_1 e^{\beta}_1 \partial_\alpha X^i\partial_\beta X^j+ e^\alpha_0\lambda^i \partial_\alpha X^j)\delta_{i j}+i \delta_{i j}e^{\alpha}_0 (\psi^i\partial_\alpha\psi^j+\tilde{\psi}^i\partial_\alpha\tilde{\psi}^j)]\,,
\eag
\ee
where $e^\alpha_A$ is the world-sheet tetrad  with $e=\det(e^A_\alpha)$. The first line of Eq. \eqref{action} appeared as the action of the $\mathcal{N}=2$ homogeneous tensionless superstring \cite{Chen:2023esw, Bagchi:2018wsn}, which considered the holomorphic and antiholomorphic fermions rescaled in the same way in taking the tensionless limit. The bosonic part of Carrollian string theory was introduced in \cite{Gomis:2023eav}, where the spacetime coordinates in the Poincar\'e directions and the Carrollian directions rescale differently in taking the tensionless limit. T duality has been demonstrated in the Carrollian strings at the level of the string sigma model without the fermionic part\cite{Gomis:2023eav}.

In the flat gauge $e^\alpha_A=\delta^\alpha_A$ \cite{Chen:2023esw},
the the action becomes\footnote{In our paper, the product denoted by the symbol $"\cdot"$ is defined to be $X\cdot Y\equiv \eta_{\mu\nu} X^\mu X^\nu=\eta_{a b}X^a Y^b+\delta_{i j} X^i Y^j\,.$}
\be\label{action}
S=\frac{1}{2\pi}\int d^2 \sigma [(\partial_0 X^a)^2+(-\partial_1 X^i \partial_1 X^i+\lambda^i \partial_0 X^i)+i (\psi\cdot\partial_0\psi+\tilde{\psi}\cdot\partial_0\tilde{\psi})]\,.
\ee
The gauge-fixed action is invariant under the world-sheet translations and the following world-sheet supersymmetric transformations,
\be\label{susyTran1}
\bag
&\delta_{\epsilon} X^a=i\epsilon \psi^a\,,\quad\delta_{\epsilon} \psi^a=-\epsilon \partial_0 X^a\,,\quad\delta_{\epsilon} \tilde{\psi}^a=\delta_{\epsilon} \tilde{\psi}^i=0,\\
&\delta_{\epsilon} X^i=0\,,\quad\delta_{\epsilon} \lambda^i=2i\epsilon\partial_1\psi^i\,,\quad\delta_{\epsilon} \psi^i=-\epsilon \partial_1 X^i\,,
\eag
\ee
and
\be\label{susyTran2}
\bag
&\delta_{\tilde{\epsilon}} X^a=i\tilde{\epsilon} \tilde{\psi}^a\,,\quad\delta_{\tilde{\epsilon}} \tilde{\psi}^a=-\tilde{\epsilon} \partial_0 X^a\,,\quad\delta_{\tilde{\epsilon}} \psi^a=\delta_{\tilde{\epsilon}} \psi^i=0,\\
&\delta_{\tilde{\epsilon}} X^i=0\,,\quad\delta_{\tilde{\epsilon}} \lambda^i=2i\tilde{\epsilon}\partial_1\tilde{\psi}^i\,,\quad\delta_{\tilde{\epsilon}} \tilde{\psi}^i=-\tilde{\epsilon} \partial_1 X^i\,.
\eag
\ee
Under this gauge, the full spacetime transformation is given by
\begin{equation}
    \begin{aligned}
        &\delta_\Lambda X^a=\Lambda^a_b X^b+\Lambda^a_i X^i\,,\quad \delta_\Lambda X^i=\Lambda^i_j X^j\,,\quad\delta_\Lambda\lambda^i=-2\Lambda^a_i\partial_0 X_a\,,\quad\\
        &\delta_\Lambda\psi^a=\Lambda^a_b \psi^b\,,\quad \delta_\Lambda\psi^i=\Lambda^i_j \psi^j\,,
    \end{aligned}
\end{equation}
where the infinitesimal transformation parameters $\Lambda^a_b$, $\Lambda^i_j$ come from the symmetry groups $SO(1,9-q)$, $SO(q)$ and $\Lambda^a_i$ represent the Carroll-like boost transformation. It can be checked that the spacetime transformation commutes with the world-sheet SUSY, $[\delta_\Lambda,\delta_\epsilon]=0$.

The equations of motion are given by
\be
\partial_0^2 X^a=0\,,\quad\partial_0 X^i=0\,,\quad 2\partial_1^2 x^i=\partial_0 \lambda^i\,,\quad \partial_0 \psi^\mu=\partial_0 \tilde{\psi}^\mu=0\,.
\ee
The physical constraints are given by varying the tetrad fields $e^\alpha_A$ in the action (\ref{actionCo}); By the equations of motion, they are either zero or related to the stress tensor $T$ and $M$ in the BMS field theory
\be
\bag
&\frac{\delta S}{\delta e^0_1}=0\,,\quad\frac{\delta S}{\delta e^1_0}=-\frac{1}{2\pi}T\,,\\
&\frac{\delta S}{\delta e^1_1}=\frac{\delta S}{\delta e^0_0}=\frac{1}{2\pi}M\,.
\eag
\ee
Note that the string action (\ref{actionCo}) was already under the partial gauge such that the world-sheet gravitino fields vanish. The form of the action before partially gauging can be found in \cite{Chen:2023esw}, where the variation of the world-sheet gravitino fields leads to the constraints corresponding to the supercurrents. Here the supercurrents can be derived by using the Noether's theorem with respect to the residual supersymmetry shown in (\ref{susyTran1}) and (\ref{susyTran2}).

One may also consider $\mathcal{N}=1,\mathcal{N}=3,\mathcal{N}=4$ homogeneous Carrollian superstrings. In these cases, by requiring the vanishing of the quantum anomaly the critical dimensions would formally be $D=\frac{82}{5}, D=\frac{38}{7}, D=2$ rather than $D=10$ for the $\mathcal{N}=2$ case \cite{Chen:2023esw}. The requirement with noninteger critical dimensions can not be satisfied within our constructions. In this work, we would like to focus on the $\mathcal{N}=2$ case because its structure is more like the usual tensile type II superstring.

\subsection{T duality}

In this subsection, for simplicity, to show the T duality among the Carrollian string theories with different $q$,  we consider a Carrollian string theory with one compactified spatial direction $X^a$ with radius $R_\text{T}$, which could be T dual to another Carrollian string theory with one compactified Carrollian spatial direction $X^i$ with $R_\text{C}$. These two Carrollian string theories are T dual to each other, i.e., their excitations are the same by swapping the zero modes for the compactified bosonic coordinates. In the following discussions in this subsection, we denote the compactified $X^a$ and $X^i$ as $X_{\text{T}}$  and  $X_{\text{C}}$ respectively, and at the same time denote $(\psi^a,\tilde{\psi}^a)$ and $(\psi^i,\tilde{\psi}^i)$ as $(\psi_{\text{T}},\tilde{\psi}_{\text{T}})$ and $(\psi_{\text{C}},\tilde{\psi}_{\text{C}}$).

Let us start with the action of the compactified $X_{\text{T}}$
\be
S=\frac{1}{2\pi}\int d^2 \sigma [(\partial_0 X_{\text{T}})^2+i(\psi_{\text{T}}\partial_0\psi_{\text{T}}+\tilde{\psi}_{\text{T}}\partial_0\tilde{\psi}_{\text{T}})]\,,
\ee
where the bosonic part is a two-dimensional (2D) electric Carrollian scalar theory on the world sheet\cite{Hao:2021urq}.
Solving its equations of motion, we have the mode expansions
\be
X_{\text{T}}=x+\frac{1}{2}p\tau+w_\text{T}R_\text{T}  \sigma+\frac{i}{2}\sum_{n\neq0}\frac{1}{n}(A_n-i n \tau B_n)e^{-i n \sigma}\,,\quad p=\frac{k_\text{T}}{R_\text{T}}\,,
\ee
and
\be
\psi_{\text{T}}=\sum_r \psi_r e^{- i r \sigma}\,,\hs{3ex}
\tilde{\psi}_{\text{T}}=\sum_r \tilde{\psi}_r e^{- i r \sigma}\,.
\ee
The derivative of $X_{\text{T}}$ is
\be
\partial_1 X_{\text{T}}=\frac{1}{2} \sum_{n} A_n e^{-i n\sigma}\,,\hspace{3ex}
\partial_0 X_{\text{T}}=\frac{1}{2} \sum_{n} B_n e^{-i n\sigma}\,,
\ee
where
\be\label{zeroMTS}
A_0\equiv 2w_\text{T}R_\text{T} \,,\quad B_0\equiv k_\text{T}/R_\text{T}\,.
\ee
The stress tensor is
\be
T=-2\partial_0 X_{\text{T}} \partial_1 X_{\text{T}}-i\psi_{\text{T}}\partial_1\psi_{\text{T}}-i\tilde{\psi}_{\text{T}}\partial_1\tilde{\psi}_{\text{T}}\,,\hs{3ex}
M=-\partial_0 X_{\text{T}} \partial_0 X_{\text{T}}\,.
\ee
The supercurrent is
\be
j=-2\psi_{\text{T}}\partial_0 X_{\text{T}}\,,\hs{3ex}
\tilde{j}=-2\tilde{\psi}_{\text{T}}\partial_0 X_{\text{T}}\,.
\ee

Next, consider the excitations in $X_{\text{C}}$. Now the action for $X_{\text{C}}$ is a 2D magnetic Carrollian scalar theory on the world sheet,
\be
S=\frac{1}{2\pi}\int d^2\sigma[(-\partial_1 X_{\text{C}} \partial_1 X_{\text{C}}+\lambda \partial_0 X_{\text{C}})+i(\psi_{\text{C}}\partial_0\psi_{\text{C}}+\tilde{\psi}_{\text{C}}\partial_0\tilde{\psi}_{\text{C}})]\,.
\ee
Solving its equations of motion, we have the mode expansions
\be
\bag
X_{\text{C}}&=x+w_\text{C}R_\text{C}  \sigma+\frac{i}{2}\sum_{n\neq0}\frac{1}{n}B_n e^{-i n \sigma}\,,\\
\lambda&=p+\sum_{n\neq 0} (A_n-i n \tau B_n)e^{-i n \sigma}\,,\quad p=\frac{k_\text{C}}{R_\text{C}}\,,\\
\psi_{\text{C}}&=\sum_r \psi_r e^{- i r \sigma}\,,\\
\tilde{\psi}_{\text{C}}&=\sum_r \tilde{\psi}_r e^{- i r \sigma}\,.
\eag
\ee
Similarly, we can denote the zero number as
\be\label{zeroMCS}
A_0\equiv k_\text{C}/R_\text{C}\,,\quad B_0\equiv 2w_\text{C}  R_\text{C} \,.
\ee
The stress tensor is 
\be
T=-\lambda\partial_1 X_\text{C}-i\psi_{\text{C}}\partial_1\psi_{\text{C}}-i\tilde{\psi}_{\text{C}}\partial_1\tilde{\psi}_{\text{C}}\,,\hs{3ex}
M=-\partial_1 X_\text{C}\partial_1 X_\text{C}\,.
\ee
The supercurrent is
\be
j=-2\psi_{\text{C}}\partial_1 X_{\text{C}}\,,\hs{3ex}
\tilde{j}=-2\tilde{\psi}_{\text{C}}\partial_1 X_{\text{C}}\,.
\ee

Now, comparing the electric sigma model with the magnetic one, they have the same canonical quantization condition
\be\label{homoComm}
[x,p]=i\,,\hs{3ex}
[A_m,B_n]=2m\delta_{m+n,0}\,,\hs{3ex}
\{\psi_r,\psi_s\}=\frac{1}{2}\delta_{r+s}\,,\hs{3ex}
\{\tilde{\psi}_r,\tilde{\psi}_s\}=\frac{1}{2}\delta_{r+s}\,,\hs{3ex}
\ee
and the same super-$\mathfrak{bms}_3$ generators from the stress tensor and supercurrent
\be\label{homoCons}
\bag
&L_n=-\frac{1}{2\pi}\int d\sigma (T+i n M)e^{i n \sigma}=\frac{1}{2}\sum_m :A_m B_{n-m}:+\sum_r (r-\frac{1}{2}n)(:\psi_r\psi_{n-r}+\tilde{\psi}_r\tilde{\psi}_{n-r}:)\,,\\
&M_n=-\frac{1}{2\pi}\int d\sigma M e^{i n \sigma}=\frac{1}{4}\sum_m B_m B_{n-m}\,,\\
&H_r=-\frac{1}{2\pi}\int d\sigma j e^{i n \sigma}=\sum_m B_m \psi_{r-m}\,,\\
&\tilde{H}_r=-\frac{1}{2\pi}\int d\sigma \tilde{j} e^{i n \sigma}=\sum_m B_m \tilde{\psi}_{r-m}\,,\\
\eag
\ee
except for the bosonic zero modes $A_0, B_0$. One can verify the super-$\mathfrak{bms}_3$ algebra (\ref{bmsalgebra}) by using (\ref{homoCons}) and the commutation relations (\ref{homoComm}) \cite{Bagchi:2018wsn}. Notice that the zero modes in the two theories can be identified after a T duality transformation
\be
w\longleftrightarrow k\,,\quad R\longleftrightarrow\frac{1}{2R}\,.
\ee
Thus, two different theories are equivalent.

It is subtle to discuss the T duality symmetry among the supersymmetric theories in flat backgrounds. We can classify Carrollian superstring in the flat background as type A or type B by the chirality of the zero-mode $\psi_0^a$ and $\tilde{\psi}_0^a$, see Eq.(\ref{chirality}). However, the T duality here does not simply map type A(B) to type B(A) or map type A(B) to type A(B), because when $q$ increases, the original spacetime spin group, a double covering of $SO(1,9-q)$, gets modified. This is weird, in contrast to the usual tensile superstring theory. Although there is no direct relation between the chirality of the theories with different $q$, the T duality is clear in the finite radius case.

\subsection{Spectrum}\label{secSpectrum}
Next, we want to discuss the spectrum of a Carrollian string theory in a flat background. Let us first consider the case without compactified spatial direction.
The mode expansions are
\be\label{modeExpan}
\bag
&X^a=x^a+\frac{1}{2}A_0^a\sigma+\frac{1}{2}B^a_0\tau+\frac{i}{2}\sum_{n\neq0}\frac{1}{n}(A^a_n-i n\tau B^a_n) e^{-i n\sigma}\,,\\
&X^i=x^i+\frac{1}{2}B^i_0 \sigma+\frac{i}{2}\sum_{n\neq 0}\frac{1}{n}B^i_n e^{-i n\sigma}\,,\\
&\lambda^i=\sum_{n}(A^i_n-i n\tau B^i_n) e^{-i n\sigma}\,,\\
&\psi^\mu=\sum_r \psi_r^\mu e^{-i r\sigma}\,,\\
&\tilde{\psi}^\mu=\sum_r \tilde{\psi}_r^\mu e^{-i r\sigma}\,,
\eag
\ee
where the mode number $r$ is a half-integer for the NS sector and an integer for the R sector. The modes satisfy the commutation relations in the canonical quantization as
\be\label{commutator}
\bag
[x^\mu,p^\nu]=i \eta^{\mu\nu}\,,&\qquad
[A^\mu_m, B^\nu_n]=2m\delta_{m+n}\eta^{\mu\nu}\,,\\
\{\psi^\mu_r,\psi^\nu_s\}=\frac{1}{2}\delta_{r+s}\eta^{\mu\nu}\,,&\qquad
\{\tilde{\psi}^\mu_r,\tilde{\psi}^\nu_s\}=\frac{1}{2}\delta_{r+s}\eta^{\mu\nu}\,.
\eag
\ee
The flipped vacuum \cite{Bagchi:2020fpr, Bagchi:2018wsn} is defined by the highest-weight representation
\be
\bag
A_n^\mu|0\rangle=B_n^\mu|0\rangle=0\qquad&n>0\,,\\
\psi_r^\mu|0\rangle=\tilde{\psi}_r^\mu|0\rangle=0\qquad&r>0\,.\\
\eag
\ee

To obtain the spectrum of physical states, we need an analog to the Virosoro constraint.
The stress tensors and supercurrents can be rewritten in terms of the modes
\be
\bag
&T=-2\partial_0 X_a\partial_1 X^a-\lambda^i\partial_1 X^i-i\psi\cdot\partial_1\psi-i\tilde{\psi}\cdot\partial_1\tilde{\psi}\,,\\
&M=-(\partial_0 X^a)^2-(\partial_1 X^i)^2\,,\\
&j=-2\psi_a\partial_0 X^a-2\psi^i\partial_1 X^i\,,\\
&\tilde{j}=-2\tilde{\psi}_a\partial_0 X^a-2\tilde{\psi}^i\partial_1 X^i\,,\\
\eag
\ee
with super-$\mathfrak{bms}_3$ generators $L_n$, $M_n$, $H_r$ and $\tilde{H}_r$ being
\be\label{superBMSgenerators}
\bag
&L_n=\frac{1}{2}\sum_m :A_{m}\cdot B_{n-m}:+\sum_r(r-\frac{1}{2}n)(:\psi_{r}\cdot\psi_{n-r}+\tilde{\psi}_{r}\cdot\tilde{\psi}_{n-r}:)\,,\\
&M_n=\frac{1}{4}\sum_m B_{m}\cdot B_{n-m}\,,\\
&H_r=\sum_m B_m\cdot\psi_{r-m}\,,\quad \tilde{H}_r=\sum_m B_m\cdot\tilde{\psi}_{r-m}\,.\\
\eag
\ee
Specifically, 
\be
\bag
&L_0=\frac{1}{2}A_{0}\cdot B_{0}+N_b+N_f+\tilde{N}_f\,,\\
&M_0=\frac{1}{4}B_0^2+\frac{1}{4}\sum_{m\neq0} B_m\cdot B_{-m}\,,
\eag
\ee
where $N$'s are number operators,
\be
\bag
N_b=\frac{1}{2}\sum_{m\neq0} :A_{-m}\cdot B_{m}:\,,\\
N_f=2\sum_{r>0} r\psi_{-r}\cdot\psi_{r}\,,\\
\tilde{N}_f=2\sum_{r>0} r\tilde{\psi}_{-r}\cdot\tilde{\psi}_{r}\,.\\
\eag
\ee

We require that the string physical states $|\text{phys}\rangle$ satisfy the highest-weight condition
\be
(L_n-a_L \delta_{n,0})|\text{phys}\rangle =M_n|\text{phys}\rangle=H_r|\text{phys}\rangle=H_r|\text{phys}\rangle=0 \quad\text{for}\quad n\geq0, r\geq0\,,
\ee
where $a_L$ is the normal-ordering constant which can be found in \cite{Chen:2023esw}. From the commutating relation between $A_m$ and $B_n$ in Eq.(\ref{commutator}), we notice that the operator $\sum_{m>0}B_{-m}\cdot B_{m}$ in $M_0$ acting on $A_{-n}$ would turn it to $B_{-n}$. Thus any eigenvector of $M_0$ in the bosonic excitations must be of the form $B_{-n_1}^{\mu_1}B_{-n_2}^{\mu_2}...|p\rangle$, which has zero norm except for the zero-mode state $|p\rangle$. A physical state with zero norm has severe problems when one tries to normalize the field it is associated with, and its 2-point function vanishes universally, as we will show in the next section.  
From another perspective, the bosonic excitations form BMS multiplets, see (\ref{multiplet}), which do not meet the requirement that the associated vertex operator must be a BMS singlet, see (\ref{integratedop}).  Therefore, we do not consider the bosonic excitations in the following discussion.

One may consider inhomogeneous superstrings as well (see appendix \ref{appendixC} for details) in which the fermionic sector is of magnetic type\cite{Bagchi:2017cte, Yu:2022bcp, Hao:2022xhq, Banerjee:2022ocj}. In this case, there would be analogous fermionic terms appearing in the $M_0$ expressions, see (\ref{M0express}). Consequently, all excitations would have a null norm, and only the zero modes would survive, which is not what we want.
Thus, in this work, we focus on the homogenous Carrollian superstring, where all bosonic excitations have zero norms while all fermionic excitations are normalizable. Therefore, we may ignore the bosonic excitations and only study the fermionic excitations.

We take the flat direction as the large $R$ limit of the compactified spatial direction of radius $R^a_{\text{T}}$ or $R^i_{\text{C}}$. Before taking the large $R$ limit, the zero modes of the string are given by
\be
\bag
&A_0^a=2w^a_{\text{T}}R_{\text{T}}\,,\quad B_0^a=k^a_{\text{T}}/R_{\text{T}}\,,\\
&A_0^i=k^i_{\text{C}}/R_{\text{C}}\,,\quad B_0^a=2w^i_{\text{C}}/R_{\text{C}}\,.
\eag
\ee
If we turn all winding numbers to zero  in the large $R$ limit  naively, which is assumed in the usual tensile string theory, the zero mode of the string becomes
\be
\bag
&A_0^a=0\,,\quad B_0^a=p^a\,,\\
&A_0^i=p^i\,,\quad B_0^a=0\,.
\eag
\ee
The highest-weight condition of $M_0$  determines the mass condition $(p^a)^2=0$, while the one of $L_0$ gives the conditions,
\be
\bag
&N_f+\tilde{N}_f=1\quad\text{for NS-NS sector}\,,\\
&N_f+\tilde{N}_f=\frac{1}{2}\quad\text{for NS-R sector}\,,\\
&N_f+\tilde{N}_f=\frac{1}{2}\quad\text{for R-NS sector}\,,\\
&N_f+\tilde{N}_f=1\quad\text{for R-R sector}\,,\\
\eag
\ee
which means that the spectrum must be truncated.

Furthermore, in order to obtain spacetime SUSY based on these physical conditions, we still need to impose an independent GSO projection, which is realized by defining the $G$ parity operator as
\be
\bag
&G=(-)^{1+2\sum_{r>0} \psi_{-r}\psi{r}} &\quad\text{for NS sector}\,,\\
&G=\left(\prod_{\mu}\psi_0^\mu\right)(-)^{2\sum_{r>0} \psi_{-r}\psi_{r}} &\quad\text{for R sector}\,,\\
\eag
\ee
and the same for $\tilde{G}$. Then we require the physical states surviving the GSO projection to have positive $G$ parity and $\tilde{G}$ parity. 
Finally, we find that the truncated spectrum is given by
\be
\bag
&|\text{NS-NS}\rangle=f_{\mu\nu}\psi^\mu_{-\frac{1}{2}}\tilde{\psi}^\nu_{-\frac{1}{2}}|p^\mu\rangle\,,\\
&|\text{NS-R}\rangle=a_{\mu}^{\tilde{I}} \psi^\mu_{-\frac{1}{2}}|p^\mu,\tilde{I}\rangle\,,\\
&|\text{R-NS}\rangle=b_{\nu}^I\tilde{\psi}^\nu_{-\frac{1}{2}}|p^\mu,I\rangle\,,\\
&|\text{R-R}\rangle=c^{I \tilde{J}}|p^\mu,I,\tilde{J}\rangle\,,\\
\eag
\ee
where the index $I$ and $\tilde{I}$ denote the spin spaces as the representation of the fermionic zero modes $(\psi_0^{\mu})_{I J}$ and $(\tilde{\psi}_0^{\mu})_{\tilde{I} \tilde{J}}$.
The coefficients in these excitations should satisfy
\be
\bag
&\qquad\qquad p^a f_{a \mu}=p^a f_{\mu a}=p^a a_a^{\tilde{I}}=p^a b_a^{I}=0\,,\\
&p_a (\tilde{\psi}_0^a)_{\tilde{I} \tilde{J}} a_{\mu}^{\tilde{J}}=p_a (\psi_0^a)_{I J} b_{\mu}^{J}=p_a (\tilde{\psi}_0^a)_{\tilde{J} \tilde{K}} c^{I \tilde{K}}=p_a (\psi_0^a)_{I K} c^{K \tilde{J}}=0\,.
\eag
\ee
As in the usual tensile superstring theory, we can classify the Carrollian string theory as type A and type B. First, we define the chirality operator as
\be\label{chirality}
\Gamma^{10-q}_{I J}=(\psi^0 \psi^1 ... \psi^{9-q})_{I J}\,,\quad \tilde{\Gamma}^{10-q}_{\tilde{I} \tilde{J}}=(\tilde{\psi}^0 \tilde{\psi}^1 ... \tilde{\psi}^{9-q})_{\tilde{I} \tilde{J}}\,,
\ee
then we refer the theory to be of type $A(B)$ if the chiralities of the $\psi$ and the $\tilde{\psi}$ parts are different(the same).
Here, the total number of spacetime bosonic excitations matches that of fermionic excitations, which implies the spacetime supersymmetry. The spacetime supersymmetry for the tensionless superstring in the GS formalism is shown in \cite{Lindstrom:1993yb}.

However, this is not the whole story because the winding number $w^a_{\text{T}}$ can generally be nonzero.
Note that the winding numbers $w^a_{\text{T}}$ and $w^i_{\text{C}}$ appear in $A_0^a$ and $B_0^i$ respectively, and $M_0$ only contains $B_0$, thus the condition $M_0|\text{phys}\rangle=0$  only requires $w^i_{\text{C}}=0$ after taking the large $R$ limit for the compactified directions.
Taking the potential nonzero winding number $w^a_\text{T}$ into account, and considering  the large $R_\text{T}^a$ limit and large $k_\text{T}^a$ at the same time so as to have a continuous-valued momentum $p^a$

\be
\frac{k_\text{T}^a}{R^a_\text{T}}\rightarrow p^a\,,
\ee
we find that the physical conditions become $(p^a)^2=0$ and
\be\label{phyCond}
N_f+\tilde{N}_f-a_L=-w_\text{T}^a k_\text{T}^a\rightarrow -w_\text{T}^a p^a R_{\text{T}}^a\rightarrow +\infty\,.
\ee
We see that for a Carrollian string theory except for $q=9$, the spectrum with zero mass is composed of an infinite number of states, which has the potential to form a higher spin theory. And there is a selection rule from the positivity of the number operators $N_f$ and $\tilde{N}_f$, leading to
\be
w_\text{T}^a p^a<0\,.
\ee
On the Carrollian spatial directions, the momentum $p^i$ is missing in the physical conditions. In other words, there is no $p^i$ dependence in the spectrum. This reflects the ultralocal property of Carrollian physics.

\section{Vertex operators and scattering amplitudes}\label{section3}

In this section, we construct the vertex operators associated with the physical states with zero winding, shown explicitly in the previous section, and use them to discuss the scattering amplitude. We will pay special attention to the amplitudes of gravitational modes. 

Here we would like to conformally transform the null cylinder to a null plane by
\be
(\tau,\sigma)\longrightarrow(y,x)=(i\tau e^{i \sigma},e^{i \sigma})\,,
\ee
for the convenience of the calculation. The necessary techniques in studying the 2D BMS field theory are enclosed in \cite{Hao:2021urq}.
The form of the action is invariant under the transformation, and the mode expansions become
\be
\bag
&X^a=x^a+\frac{i}{2}\sum_{n}\frac{1}{n}(A^\mu_n-n \frac{y}{x} B^\mu_n) x^{-n}\,,\\
&X^i=x^i+\frac{i}{2}\sum_{n}\frac{1}{n}B^i_n x^{-n}\,,\\
&\lambda^i=-i\sum_{n}(A^i_n-(n+1)\frac{y}{x}B^i_n) x^{-(n+1)}\,,\\
&\psi^\mu=\sum_r \psi_r^\mu x^{-(r+\frac{1}{2})}\,,\\
&\tilde{\psi}^\mu=\sum_r \tilde{\psi}_r^\mu x^{-(r+\frac{1}{2})}\,,
\eag
\ee
and the stress tensors and the supercurrents can be expanded into the form
\be
\bag
&T=-2\partial_y X_a\partial_x X^a-\lambda^i\partial_x X^i-\psi\cdot\partial_x\psi-\tilde{\psi}\cdot\partial_x\tilde{\psi}=\sum_n L_n x^{-n-2}-\sum_n (n+1)y M_{n-1}x^{-n-2}\,,\\
&M=-(\partial_y X^a)^2-(\partial_x X^i)^2=\sum_n M_n x^{-n-2}\,,\\
&j=2i(\psi_a\partial_y X^a+\psi^i\partial_x X^i)=\sum_r H_r x^{-r-\frac{3}{2}}\,,\\
&\tilde{j}=2i(\tilde{\psi}_a\partial_y X^a+\tilde{\psi}^i\partial_x X^i)=\sum_r \tilde{H}_r x^{-r-\frac{3}{2}}\,,\\
\eag
\ee
where the generators $L_n, M_n, H_r, \tilde{H}_r$ are the same as in (\ref{superBMSgenerators}) and satisfy homogeneous super-$\mathfrak{bms}_3$ algebra (\ref{bmsalgebra}).
It is not hard to calculate the operator-product expansions (OPEs) among them. The nonzero OPEs are shown as
\be\label{freeOPE}
\bag
&X^i(y,x)X^j(0,0)\sim 0\,,\quad X^i(y,x) \lambda^j(0,0)\sim \frac{1}{x}\delta_{i j}\,,\quad
\lambda^i(y,x) \lambda^j(0,0)\sim \frac{4y}{x^3}\delta_{i j}\,,\\
&X^a(y,x) X^b(0,0)\sim -\frac{y}{2x}\eta^{a b}\,,\quad\psi^\mu(y,x) \psi^\nu(0,0)\sim \frac{1}{2x}\eta^{\mu\nu}\,,\quad\tilde{\psi}^\mu(y,x) \tilde{\psi}^\nu(0,0)\sim \frac{1}{2x}\eta^{\mu\nu}\,,
\eag
\ee
where the fields $X^i$, $\psi^\mu$, and $\tilde{\psi}$ are all primary BMS singlets,
\be
\bag
&T(y,x)X^i(0,0)\sim\frac{\partial_x X^i}{x}\,,\quad M(y,x)X^i(0,0)\sim 0\,,\\
&T(y,x)\psi^\mu(0,0)\sim\frac{\partial_x \psi^i}{2x}\,,\quad M(y,x)\psi^i(0,0)\sim 0\,,\\
&T(y,x)\tilde{\psi}^\mu(0,0)\sim\frac{\partial_x \tilde{\psi}^i}{2x}\,,\quad M(y,x)\tilde{\psi}^i(0,0)\sim 0\,,\\
\eag
\ee
while $(\partial_x X^i,\frac{\lambda^i}{2})$ and $(\partial_y X^a, \partial_x X^a)$ are primary BMS doublets, which can be expressed as $(O_0,O_1)$ satisfying
\be
\bag
&T(y,x)O_0\sim\frac{O_0}{x^2}+\frac{\partial_x O_0}{x}\,,\quad M(y,x)O_0\sim \frac{\partial_y O_0}{x}\,,\\
&T(y,x)O_1\sim\frac{O_1}{x^2}+\frac{\partial_x O_1}{x}-\frac{2y O_0}{x^3}-\frac{y\partial_y O_1}{x^2}\,,\quad M(y,x)O_1\sim \frac{O_0}{x^2}+\frac{\partial_y O_1}{x}\,.\\
\eag
\ee
When the primary operator $\boldsymbol{O}(y,x)$ is a multiplet\footnote{The multiplet representations also exist in higher dimensional Carrollian field theory, see \cite{Chen:2021xkw,Chen:2023pqf,Saha:2023hsl}} with rank $r$, the corresponding definition should become
\be\label{multiplet}
[L_0,\boldsymbol{O}(0,0)]=\Delta \boldsymbol{O}(0,0)\,,\quad [M_0,\boldsymbol{O}(0,0)]=\boldsymbol{\xi} \boldsymbol{O}(0,0)\,,
\ee
where $\boldsymbol{O}=(O_1,O_2,...,O_r)^{T}$ and
\be
\boldsymbol{\xi}=\begin{pmatrix}\xi&&&\\1&\xi&&\\&\ddots&\ddots&\\&&1&\xi\end{pmatrix}_{r\times r}\,.
\ee

To construct all the vertex operators that appear in a general string amplitude, we need the $bc$-ghost, $\beta\gamma$-ghost fields, and picture-changing operators in the standard BRST quantization process \cite{Chen:2023esw}. We will focus on the bosonic part, which differs significantly from the standard bosonic string. We would simply skip the technical details on fermionic fields, which are quite similar to the usual case, and directly use the results to discuss tree-level amplitudes, referring to the standard process in the textbook\cite{Polchinski:1998rq, Polchinski:1998rr}. The calculation is almost the same, but the (anti)-holomorphic coordinates $z,\bar{z}$ in the Riemann surface are all replaced by the coordinate $x$ in the null plane. This is because the tensionless limit can effectively be transformed to the ultrarelativistic limit \cite{Hao:2021urq}
\be
z\rightarrow \epsilon y+x\,,\quad \bar{z}=-\epsilon y+x\,,\quad \epsilon\rightarrow 0\,,
\ee
and the homogeneous fermions rescaled in the same way when taking the ultrarelativistic limit \cite{Bagchi:2018wsn}.

The momentum eigenstate with momentum $p^\mu=(p^a, p^i)$ is constructed from the vacuum by
\be
|p\rangle=\exp{i p\cdot x}|0\rangle\,.
\ee
One can verify that it is a primary state satisfying
\be
\bag
&L_n |p\rangle=0\quad(n\geq0)\,,\\
&M_n |p\rangle=\frac{(p^a)^2}{4}\delta_{n,0}\quad(n\geq0)\,.\\
\eag
\ee
So the corresponding vertex operator $V_p(y,x)$ must satisfy $V_p(0,0)|0\rangle=|p\rangle$ and be a primary BMS operator. The vertex operator is of the form
\be
V_p(y,x)=:\exp(i p\cdot X(y,x)):\,,
\ee
satisfying
\be
\bag
&T(y,x)V_p(0,0)\sim\frac{\partial_x V_p}{x}-\frac{y\partial_y V_p}{x^2}-\frac{(p^a)^2 y V_p}{2x^3}\,,\\
&M(y,x)V_p(0,0)\sim\frac{\partial_y V_p}{x}+\frac{(p^a)^2 V_p}{4x^2}\,,
\eag
\ee
and
\be
\bag
&X^a(y,x) V_p(0,0)\sim\frac{-i p^a y}{2x}V_p\,,\\
&V_{p_1}(y,x)V_{p_2}(0,0)\sim\exp((\eta_{a b}p_1^a p_2^b)\frac{y}{2x})V_{p_1+p_2}\,.
\eag
\ee

For an integrated vertex operator appearing in the scattering amplitude, $\int d x d y V$ must be a primary singlet of dimension $(\Delta,\xi)=(0,0)$, which means that the vertex operator $V$ must be a primary singlet of dimension $(2,0)$, because under a general BMS transformation
\be
(y,x)\longrightarrow(y', x')=(y f'(x)+g(x),f(x))\,,
\ee
the integrated vertex operator transforms as
\be\label{integratedop}
\int d x d y V\longrightarrow \int d x' d y' V'=\int d x d y |f'|^2 |f'|^{-\Delta} e^{-\xi\frac{g'+y f''}{f'}} V\,.
\ee
This requirement reconciles with the physical conditions demonstrated in the previous section.

Next, we discuss the scattering amplitudes relating to the Carrollian gravity. In the bosonic string case, the gravitational and dilational modes are from the excitations
\be
|f,p\rangle=f_{\mu\nu}B^\mu_{-1} B^\nu_{-1} |p\rangle\,,\quad f_{\mu\nu}=f_{\nu\mu}\,,f_{a \mu}p^{a}=0\,.
\ee
They all have zero norm, meaning their associated vertex operators
\be
V_{f,p}=:(f_{a b}\partial_y X^a \partial_y X^b+2f_{a i}\partial_y X^a \partial_x X^i+f_{i j}\partial_x X^i \partial_x X^j)\exp(i p\cdot X):\,
\ee
always have vanishing two-point functions,
\be
\langle V_{f_1,p_1}(y_1,x_1)V_{f_2,p_2}(y_2,x_2)\rangle\propto f_{1 a b}p_2^a p_2^b f_{2 c d}p_1^c p_1^d \delta(p_1+p_2)/x_{12}^4=0\,.
\ee
The general tree-level $n$-point  amplitudes among them are given by
\be
\mathcal{A}_n\propto g_s^{n-2}x_{12}^2 x_{23}^2 x_{31}^2 \prod_{I=4}^n \int d x_I d y_I \langle V_1...V_I(y_I,x_I)...V_n\rangle\,,
\ee
where $g_s$ is the string coupling. We note that any amplitude involving $X^i$ must vanish because the vertex operators have no $\lambda^i$. This means that any excitations along the Carrollian spatial directions are nondynamical. This is not true when considering the superstring case.

In the superstring case, the vertex operator in $(-1,-1)$ picture for a physical state based on the NS-NS sector is given by $V^{(-1,-1)}=e^{-\phi-\tilde{\phi}}\mathcal{O}$, where $\phi$ and $\tilde{\phi}$ are the bosonization of $\beta\gamma$ ghost, satisfying $\phi(y,x)\phi(0,0)\sim-\log{x}$ and the same for $\tilde{\phi}$, and the operator $\mathcal{O}$ is given by
\be
\mathcal{O}=:f_{\mu\nu}\psi^\mu\tilde{\psi}^\nu\exp{i p\cdot X}:\,,
\ee
satisfying
\be
(p^a)^2=0\,,\quad f_{\mu a}p^a=f_{a \mu}p^a=0\,.
\ee
The corresponding vertex operator in $(0,0)$ picture is $V^{(0,0)}=H_{-\frac{1}{2}}\tilde{H}_{-\frac{1}{2}}\cdot \mathcal{O}$, where $H_{-\frac{1}{2}}$ and $\tilde{H}_{-\frac{1}{2}}$ are given by (\ref{superBMSgenerators}), and its explicit expression is
\be
V^{(0,0)}_{f,p}=:f_{\mu \nu}(-i\partial' X^\mu+p^a\psi^a\psi^\mu)(-i\partial' X^\nu+p^a\tilde{\psi}^a\tilde{\psi}^\nu)\exp{i p\cdot X}:\,,
\ee
where $\partial'X^\mu\equiv \partial_y X^\mu\eta(\mu)-\partial_x X^\mu(1-\eta(\mu))$ with the $\eta$ function being defined by
\be
\eta(\mu)\equiv \left\{  \ba{ll}
1, &\mbox{when $\mu=a$,}\\
0, & \mbox{when $\mu=i$.}\ea\right. 
\ee
The general tree-level $n$-point  amplitudes are now given by
\be
\mathcal{A}_n\propto g_s^{n-2}x_{12}^2 x_{23}^2 x_{31}^2 \left(\prod_{I=4}^n \int d x_I d y_I\right) \langle V_1^{(-1,-1)}...V_I^{(-1,-1)}(y_I,x_I)...V^{(-1,-1)}_n\chi^{n-2}\rangle\,,
\ee
where there are $(n-2)$ picture-changing operators $\chi$ which can turn $V^{(-1,-1)}$ to $V^{(0,0)}$.
Especially, the three-point amplitudes are\footnote{Note that in Minkowski signature for $\eta^{\mu\nu}$, the general 3-pt amplitudes are vanishing due to kinematic reasons, thus the results here are in fact for the general signature case.}
\be\label{3ptAmp}
\bag
\mathcal{A}_3&\propto g_s x_{12}^2 x_{23}^2 x_{31}^2 \langle V_1^{(-1,-1)}V_2^{(-1,-1)}V^{(0,0)}_3\rangle\\
&=
\frac{i g_s}{64} (2\pi)^{10} \delta^{(10)}(p_1+p_2+p_3)f_{1\mu\sigma}f_{2\nu\omega}
f_{3\rho\lambda}V^{\mu\nu\rho}V^{\sigma\omega\lambda}\,,
\eag
\ee
where
\be
V^{\mu\nu\rho}=\eta^{\mu\nu}p_{1 2}^{\rho}\eta(\rho)+\eta^{\nu\rho}p_{2 3}^{\mu}\eta(\mu)+\eta^{\rho\mu}p_{3 1}^{\nu}\eta(\nu)\,.
\ee
The result is the same as the usual tensile superstring case except for the appearance of $\eta$ functions.
\begin{itemize}
    \item When $q=0$, we return to the  tensionless string theory with $10D$ Minkowski target spacetime. In this case, the amplitudes are identical to those in usual type-II string theory, then the action for the gravitational modes (requiring $f_{\mu\nu}$ to be symmetric and traceless) is expected to be that of Einstein gravity.
    \item When $q=9$, which is the conventional Carrollian case, the physical condition tells $(p^0)^2=0$, that is $p^0=0$, and the amplitudes become zero. This reflects the fact that there is no local propagating gravitational mode in conventional Carrollian gravity theory\cite{Hansen:2021fxi, deBoer:2023fnj, Tadros:2023teq}.
    \item When $0<q<9$, the physical condition (dispersion relation)  is
\be
(p^a)^2=0\nRightarrow p^a=0\qquad(a=0,1,..,9-q)\,,
\ee
There could exist local propagating gravitons. Thus it would be fascinating to see the right gravity theory in the generalized Carrollian spacetime. And the result we obtained in this paper would be a guideline for constructing that theory.
\end{itemize}

\section{Conclusion and discussions}\label{section4}

In this work, we studied the closed Carrollian superstring theory in the flipped vacuum. We showed explicitly that two homogeneous Carrollian superstring theories, one with compactified Poincar\'e direction and the other with compactified Carrollian direction, share the same super-$\mathfrak{bms}_3$ algebra, and their zero modes can be identified under the usual T duality transformation. Moreover, we investigate the spectrum of a general Carrollian superstring in the flipped vacuum. One remarkable feature is that the spectrum of zero mass is not truncated and could be composed of an infinite number of states\footnote{Some discussions about tensionless bosonic string can be found in \cite{Banerjee:2023ekd, Banerjee:2024fbi}. The ambitwistor string, which can be seen as the tensionless string in a different vacuum \cite{Casali:2016atr}, was discussed in \cite{Casali:2017mss}.}, as the winding number is allowed to be nonvanishing in the infinite radius limit. We would like to emphasize that it is necessary to introduce homogenous world-sheet fermions because all the bosonic excitations and inhomogeneous fermionic excitations have zero norms \cite{Chen:2023esw}. As a result, the number of spacetime bosonic excitations matches that of fermionic excitations, indicating the spacetime supersymmetry. We furthermore constructed the vertex operators of gravitons and discussed their scattering amplitudes. We found that the form of 3-point amplitudes is reminiscent of the usual tensile superstring except for the appearance of $\eta$ functions, which reflects the ultralocal nature of Carrollian physics.

As we have shown, the Carrollian superstring theory spectrum is highly degenerate. 
If we consider a Carrollian string theory with a compactified spatial direction with radius $R_\text{T}$ or $R_\text{C}=1/(2R_{\text{T}})$, given a specific $w_{\text{C}}=k_{\text{T}}$, all the states with arbitrary $w_{\text{T}}=k_{\text{C}}$ would have the same on-shell condition
\be
m^2\equiv-(p^a)^2=(k_{\text{T}}/R_\text{T})^2=(2w_{\text{C}}R_\text{C})^2\,.
\ee
If we consider a Carrollian string theory in a flat background, from (\ref{phyCond}), we can see that the spectrum is still highly degenerate, it may have infinitely many higher-spin components. 
How to describe these highly degenerate states is an important question. As there does not exist a string scale parameter like $\alpha'$, it makes no sense to discuss low-energy effective action by decoupling the high-energy modes. In other words, we have to discuss all these modes equally, which is a feature of higher spin theory. 

Suppose that we do not consider the nonzero winding mode and focus only on the zero-mass states in the flat background. In that case, we may study the effective action of the string theory by considering the vanishing of the beta functions, just as in the nonrelativistic case \cite{Yan:2019xsf, Gomis:2020fui}. However, in this work, we chose to study the dynamics of massless fields by using the perturbative scattering amplitudes. We found that the 3-point amplitudes for gravitational modes are identical to those of usual type-II string theory except for the appearance of $\eta$ functions. The $\eta$ functions can be deduced from the ultrarelativistic limit $c\rightarrow 0$.  Considering 
a target spacetime geometry of the following metric
\be\label{CarrGeo}
d s^2=c^2\eta_{a b} d {X^a} d {X^b}+\delta_{i j} d {X^i} d {X^j}\,,
\ee
where $c$ is the speed of light, we quantize the tensile string in this spacetime. The low-energy effective action of the (super)string is still Einstein's (super)gravity. The 3-point amplitudes for the gravitons are
\be \label{3amE}
\mathcal{A}_3\propto \delta^{(10)}(p_1+p_2+p_3)f_{1\mu\sigma}f_{2\nu\omega}f_{3\rho\lambda}V'^{\mu\nu\rho}V'^{\sigma\omega\lambda}\,,
\ee
where
\be
V'^{\mu\nu\rho}=\eta^{\mu\nu}p^\rho_{1 2}+\eta^{\nu\rho}p^\mu_{2 3}+\eta^{\rho\mu}p^\nu_{3 1}\,.
\ee
There is no $\eta(\cdot)$ function in the expressions as in Eq.(\ref{3ptAmp}). Now the mass-shell condition is
\be
\eta^{a b}p_a p_b+c^2\delta^{i j}p_i p_j=(p^a)^2+c^2 (p^i)^2=m^2 c^4\,.
\ee
Effectively, every momentum along the Carrollian spatial directions is accompanied by a speed of light $c$. Thus if we take the ultrarelativistic limit $c\rightarrow 0$  in the amplitude \eqref{3amE}, we obtain the result Eq.(\ref{3ptAmp}), where the $\eta(\cdot)$ functions emerge\footnote{Note that there is no speed-of-light dependence in the polarization vectors $f_{\mu \nu}$ due to proper rescaling between Einstein's gravity and Carrollian gravity components.}.

The higher-point amplitudes in the Carrollian string theory would be very different from the ones in the usual tensile string. Typically, the amplitudes take the form
\be
V_{p_1}V_{p_2}...V_{p_n}=\exp(\sum_{i<j}^n \eta_{a b}p^a_i p^b_j\frac{y_{i j}}{2 x_{i j}})V_{\sum_i^n p_i}\,,
\ee
which involves an exponential factor depending on $y_{ij}=y_i-y_j$. 
In the case of 3-point amplitudes, due to the momentum conservation and the on-shell condition, 
\be
2\eta_{a b}p_1^a p_2^b=(p_3^a)^2=0\,,
\ee
and the same for the other two terms, 
there is no $y$-dependent exponential factor. 
However, this is not true when considering higher-point amplitudes. The exponential structure is brand new and does not appear in the usual tensile string theory because the OPEs between vertex operators in $\text{CFT}_2$ are always in the powers of the coordinates. 
This essential difference in the higher-point amplitudes should be traced to the difference in the internal propagating modes for different string theories, while the similarity in 3-point amplitudes comes from the constraint from the spacetime symmetry. It would be interesting to investigate more carefully the structure of these amplitudes and their implications on nonlinear interactions in Carrollian gravity.

In this work, we focus on the closed Carrollian strings. It is an open issue how to study the T duality in open Carrollian or tensionless string, which may involve D-branes or S-branes\cite{Bagchi:2024qsb}. We hope to address this issue in the future.

\section*{Acknowledgments}
We thank Haowei Sun and Yu-fan Zheng very much for the valuable discussions. This research is supported in part by NSFC Grant  No. 11735001, 12275004.

\appendix

\section{$\mathcal{N}=2$ inhomogenous super-$\mathfrak{bms}_3$ algebra}\label{appendixC}

In the $\mathcal{N}=2$ inhomogeneous case, the super-$\mathfrak{bms}_3$ algebra is given by\footnote{The anomalous terms $A_L(m), A_M(m), B_G(r)$ and more details can be found in \cite{Chen:2023esw}.}
\begin{equation}
    \begin{aligned}
        &\left[L_m, L_n \right] = \left(m-n\right) L_{m+n} + \delta_{m+n}A_L(m), \qquad &&\left[L_m, M_n\right] = \left(m-n\right) M_{m+n} + \delta_{m+n}A_M(m),\\
        &\left[L_n, G_r \right] = \left(\frac{n}{2}-r\right) G_{n+r}, \qquad &&\left[M_n, G_r\right] = \left(\frac{n}{2}-r\right) H_{n+r},\\
        &\left[L_n, H_r \right] = \left(\frac{n}{2}-r\right) H_{n+r}, \qquad &&\left[M_n, H_r\right] = 0,\\
        &\left\{G_r, G_s\right\} = 2 L_{r+s} + \delta_{r+s}B_G(r), \qquad &&\left\{G_r, H_s\right\} = 2 M_{r+s}, \\
        &\left\{H_r, H_s\right\} = 0.&&
    \end{aligned}
\end{equation}
This algebra can be constructed from the inhomogeneous Carrollian superstring, of which the gauge-fixed string sigma model is
\be
\bag
S=\frac{1}{2\pi}\int d^2\sigma &[(\partial_0 X^a)^2-(\partial_1 X^i)^2+\delta_{i j}\lambda^i \partial_0 X^j\\
&+i\psi_1\cdot\partial_0\psi_0+i\psi_0\cdot\partial_0\psi_1-i\psi_0\cdot\partial_1\psi_0]\,.
\eag
\ee
The mode expansions for the bosonic part are the same as in (\ref{modeExpan}). The mode expansions for the inhomogeneous fermions are
\be
\psi_0^\mu=\sum_r \psi^\mu_r e^{-i r\sigma}\,,\quad \psi_1^\mu=\sum_r \left(\frac{1}{2}\tilde{\psi}^{\mu}_r-i r \tau \psi^\mu_r\right) e^{-i r\sigma}\,.
\ee
The commutation relations are
\be\label{inhomoComm}
[x^\mu,p^\nu]=i \eta^{\mu\nu}\,,\quad[A^\mu_m, B^\nu_n]=2m\delta_{m+n}\eta^{\mu\nu}\,,\quad \{\tilde{\psi}^\mu_r,\psi^\nu_s\}=\delta_{r+s}\eta^{\mu\nu}\,,\quad \text{Others}=0\,.
\ee
Then the inhomogeneous super-$\mathfrak{bms}_3$ algebra can be constructed as
\be\label{inhomoCons}
\bag
&L_n=\frac{1}{2}\sum_m :A_{m}\cdot B_{n-m}:+\sum_r(r-\frac{1}{2}n)(:\psi_{n-r}\cdot\tilde{\psi}_r:)\,,\\
&M_n=\frac{1}{4}\sum_m B_{m}\cdot B_{n-m}+\sum_r (r-\frac{n}{2})\psi_{n-r}\psi_{r}\,,\\
&G_r=\sum_m \psi_{r-m}\cdot A_m+\sum_m\frac{1}{2}\tilde{\psi}_{r-m}B_m\,,\\
&H_r=\sum_m \psi_{r-m}\cdot B_m\,.\\
\eag
\ee
Here we can see that imposing the physical constraint $M_0|phy\rangle=0$ where
\be\label{M0express}
M_0=\frac{1}{4}\sum_m B_{m}\cdot B_{-m}+\sum_r r\psi_{n-r}\cdot\psi_{r}\,,
\ee
requires that all the physical excitations must be of the form $B^{\mu_1}_{-n_1}B^{\mu_2}_{-n_2}...\psi^{\mu_1}_{-r_1}\psi^{\mu_2}_{-r_2}...|p\rangle$. They all have zero norms according to the commutation relation (\ref{inhomoComm}) and the definition of the flipped vacuum
\be
\bag
A_n^\mu|0\rangle=B_n^\mu|0\rangle=0\qquad&n>0\,,\\
\psi_r^\mu|0\rangle=\tilde{\psi}_r^\mu|0\rangle=0\qquad&r>0\,.\\
\eag
\ee

\bibliographystyle{utphys}
\bibliography{refs}
\end{document}